\documentclass{article}

\usepackage[margin=0.75in]{geometry}
\usepackage[utf8]{inputenc}
\usepackage[title]{appendix}
\usepackage[sort&compress,square,comma,numbers]{natbib}
\usepackage[version=4]{mhchem}
\usepackage{graphicx,subfig,float}
\usepackage{tabularx}
\usepackage{amsmath,amsthm,amssymb,commath,amsfonts}
\usepackage{booktabs,multirow}
\usepackage[normalem]{ulem}
\usepackage[colorlinks=true,bookmarksopen=true]{hyperref}
\usepackage{algpseudocode}
\usepackage{algorithm}
\usepackage{setspace}
\usepackage[normalem]{ulem}
\usepackage{todonotes}

\usepackage{dcolumn}
\usepackage{bm}

\begin{document}

\title{Mesoscopic modeling of heptane: A surface tension calculation\thanks{Approved for public release (INL/JOU-20-57228)}}

\author{
Qi Rao, Yidong Xia \thanks{Corresponding author: Yidong Xia (\url{yidong.xia@inl.gov})}, Jiaoyan Li\\
Energy and Environment Science \& Technology, Idaho National Laboratory\\Idaho Falls, Idaho, USA\\
\and
Zhen Li \thanks{Co-corresponding author: Zhen Li (\url{zli7@clemson.edu})}\\
Department of Mechanical Engineering, Clemson University\\Clemson, South Carolina, USA\\
\and
Joshua McConnell, James Sutherland\\
Department of Chemical Engineering, The University of Utah\\Salt Lake City, Utah, USA\\
}





\date{}

\maketitle

\section*{\centering Abstract}
Accurate and efficient flow models for hydrocarbons are important in the development of enhanced geotechnical engineering for energy source recovery and carbon capture \& storage in low-porosity, low-permeability rock formations. This work reports an atomistically-validated, mesoscopic model for heptane based on a many-body dissipative particle dynamics (mDPD) method. In this model, each heptane molecule is coarse-grained in one mDPD bead and the mDPD model parameters are calibrated with a rigorous approach using reference data, including experimental measurements and/or molecular dynamics (MD) simulations. Results show that this mDPD model accurately predicts the bulk pressure-density relation of heptane and surface tension. Notice that our approach can be used to calibrate the mDPD model for other hydrocarbons as well, though heptane is chosen as a representative source fluid for its abundance in source rocks. Further, our timing test indicates that the mDPD model is three orders of magnitude faster than its MD counterpart for simulations of bulk heptane in equivalent volumes. Overall, this work serves as a key prerequisite for the development of accurate and efficient mesoscale models for the flow of hydrocarbons confined in mesoporous rock formations.

\vspace{1em}
\noindent {\it Keywords:} hydrocarbon, dissipative particle dynamics, many-body, coarse graining, equation of state

\section{\label{sec:intro}Introduction}

Understanding the flow and transport properties of hydrocarbons in geological mesoscale confinement (i.e. with apertures of pore channels that range from a few nanometers ($10^{-9}$m) to a few micrometers ($10^{-6}$m)) is important in the development of enhanced geotechnical engineering for energy source recovery and carbon capture \& storage in low-porosity, low-permeability rock formations. To measure those properties experimentally, however, is challenging or not currently possible as the laboratory instrumentation for mesoscale confinement with proper geological conditions is still under development. With the ever-growing capacity of computers, modeling and simulation have been adopted as an alternative for such studies and guiding the conceptual design of experiments. To perform flow simulations, computational models are normally chosen based on their applicable ranges of scales and computing costs, e.g., molecular dynamics (MD) models for tens of thousands of hydrocarbon molecules and continuum-based computational fluid dynamics (CFD) models in the macroscopic scale. Those models, however, are found either not computationally affordable (i.e. for MD) or not theoretically fit (i.e. for CFD) for describing the physics of fluids in the mesoscale. The flow properties in the mesoscale range can be regionally dominated by thermal fluctuations and molecular diffusion in nanopores while behaving similarly to continuum media in micropores.

Dissipative particle dynamics (DPD) \cite{hoogerbrugge1992simulating,groot1997dissipative}
is a class of mesoscopic fluid flow models that can potentially fill the gap between atomistic and macroscopic models. The theoretical foundation of DPD is established on the statistical mechanics \cite{groot1997dissipative,marsh1998theoretical}. The theory, methodology, recent developments, and applications of DPD are summarized by \citet{moeendarbary2009dissipative}, \citet{liu2015dissipative} and \citet{li2017dissipative}, respectively. In the original DPD method, a cluster of fluid molecules are coarse-grained as a single bead. Accordingly, the atomistic interactions between the fluid molecules are coarse-grained by simplified bead-bead interactions. The formulation of DPD also allows more structural complexities and constraints such as bond and dihedral, e.g., to model red blood cells \cite{pan2010single,pan2010single,pan2011predicting,tang2014accelerating,blumers2017gpu}. As a result of coarse-graining, the critical timestep size in DPD models usually can be many orders of magnitude larger than their MD counterpart and thus can permit the sampling of both length and time scales equivalent to those experimentally measurable, though at the expense of neglecting the details of molecular interactions. In DPD, the modeling of  multi-phase fluid flow has been made possible with a number of variant models \cite{warren2003vapor,tiwari2006dissipative,heldele2006micro,visser2006modelling,liu2006dissipative,liu2007dissipative-jcp,liu2007dissipative-wrr}. However, despite the many favorable features of DPD, there has not been a unified way to calibrate the DPD fluid model parameters for specific types of fluids by coarse-graining from their underlying MD models. To the best of our knowledge, liquid water is so far the only fluid that was rigorously calibrated for DPD, but without sufficient detail of how the model parameters were determined in literature \cite{ghoufi2011mesoscale}. Though the DPD simulations of hydrocarbon flow were also reported in case studies such as hydrocarbon recovery in a simple channel \cite{chen2014many} and complex pore networks \cite{xia2017many,xia2020gpu}, the DPD model parameters for hydrocarbons were previously determined by only roughly fitting the pattern of their contact angles with confining wall surfaces, but without accurate calibration of key material properties such as bulk density and surface tension. In order to obtain and use accurate DPD modeling for specific hydrocarbons of interest, an atomistically-informed, systematic characterization approach for DPD is urgently required.

The objective of the major effort discussed in this work is the development and application of a general method for accurate calibration of DPD fluid models based on reference data from molecular simulations or experiments. Applying the calibration approach, we have obtained a mesoscale model for heptane (\ce{C7H16}) based on a many-body variant of the original DPD method by \citet{warren2003vapor}, namely mDPD in most literature, e.g. in \citet{li2013three}. The mDPD method is reportedly suitable to model the pressure-density behavior of realistic fluids by considering the many-body interactions \cite{pagonabarraga2001dissipative,trofimov2002thermodynamic,warren2003vapor,trofimov2005constant} and has been applied in mesoscale simulations of liquid water–vapor interface tension \cite{ghoufi2011mesoscale}, contact angle characterization \cite{li2013three}, and multi-phase, multi-fluid flow in micro-channels \cite{pan2010single,chen2011many,chen2013effective,chen2014many}. Notice that the mDPD formulation assumes an isothermal condition in which the conservation of total energy is dismissed. This indicates that a set of mDPD parameters are valid for only a specific or narrow range of temperature. To model non-isothermal flows, the current mDPD model requires additional constraint for total energy conservation \cite{espanol1997dissipative,li2014energy,ripoll1998dissipative,avalos1999dynamic}, which is though not in the scope of this work. Notice that our approach can be used to calibrate the mDPD model for other hydrocarbons as well, e.g. toluene (\ce{C7H8}), though heptane is chosen in this work as a representative source fluid for its abundance in source rocks. Our mDPD simulations of bulk heptane have been rigorously validated against MD simulations and experimental data. Numerical results show that with the parameters calibrated specifically for heptane at the temperatures of interest, the mDPD model accurately predicts the pressure-density relation and key material properties including bulk density, compressibility, and surface tension. The calibration of bulk properties is an essential prerequisite for mDPD to be further calibrated for realistic liquid-wall interactions.

The rest of this work is organized as follows. The mDPD formulation is introduced in \autoref{sec:details}. In \autoref{sec:model-calibration}, we describe a general calibration approach for the mDPD model and use heptane as a target application for the model parameterization. The validation of our mDPD simulation results is presented and discussed in \autoref{sec:results}, followed by our summary and conclusions in \autoref{sec:conclusions}.

\section{\label{sec:details}Many-body DPD formulation}

In the original DPD method, the force formulation that describes the interactions between DPD particles comprises of three components \cite{hoogerbrugge1992simulating}:
\begin{equation}
    \bm{F}_{ij}=\bm{F}^C_{ij}+\bm{F}^D_{ij}+\bm{F}^R_{ij},
\end{equation}
where $\textbf{F}^C_{ij}$, $\textbf{F}^D_{ij}$, $\textbf{F}^R_{ij}$ represent a conservative force, a dissipative force, a random force between particle \textit{i} and \textit{j}, respectively. If $\boldsymbol{r}_i$ and $\boldsymbol{v}_i$ are used to denote the position and velocity of particle $i$, respectively, the dissipative force, $\textbf{F}^D_{ij}$, and the random force, $\textbf{F}^R_{ij}$, can be expressed as
\begin{equation}
    \textbf{F}^D_{ij}  = -\gamma\omega^D(r_{ij})(\boldsymbol{\hat{r}}_{ij}\cdot\boldsymbol{v}_{ij})\boldsymbol{\hat{r}}_{ij}
\end{equation}
and
\begin{equation}
    \textbf{F}^R_{ij} = \sigma\omega^R(r_{ij})\xi_{ij}\boldsymbol{\hat{r}}_{ij},
\end{equation}
where $\boldsymbol{r}_{ij} = \boldsymbol{r}_i - \boldsymbol{r}_j$, $r_{ij} = |\boldsymbol{r}_{ij}|$, $\boldsymbol{\hat{r}}_{ij} = \boldsymbol{r}_{ij}/r_{ij}$, $\boldsymbol{v}_{ij} = \boldsymbol{v}_i - \boldsymbol{v}_j$. The amplitude $\sigma$ of the random variable $\xi_{ij}$, and the viscous dissipation coefficient $\gamma$ satisfy a fluctuation-dissipation theorem: $\sigma^2 = 2{\gamma}k_BT$ and $\omega^D(r) = (\omega^R(r_{ij}))^2$. The conservative force in the original DPD is purely repulsive, which is not sophisticated enough to model the free surface phenomena and multiphase behavior. The mDPD replaces the original repulsive conservative force term with the following form:
\begin{equation}
\label{eq:conservative-force}
\textbf{F}^C_{ij} = A\omega^C(r_{ij})\boldsymbol{\hat{r}}_{ij} + B(\bar{\rho}_i + \bar{\rho}_j)\omega_d(r_{ij})\boldsymbol{\hat{r}}_{ij},
\end{equation}
where the first term is a long-range attractive force (with negative values for coefficient $A$) and the second term is a local number density-dependent, short-range repulsive force (with positive values for coefficient $B$). The weight functions $\omega^C(r_{ij})$ and $\omega_d(r_{ij})$ are chosen as $\omega^C(r_{ij}) = 1 - r_{ij}/r_c$, $\omega_d(r_{ij}) = 1 - r_{ij}/r_d$ and $\omega^R = \omega^C$ ($r_d < r_C$). The averaged local density, $\bar{\rho}_i$, at the position of particle $i$ can be computed as $\bar{\rho}_i = \sum_{j{\neq}i}\omega_\rho(r_{ij})$, where the normalized weight function, $\omega_\rho$, needs to satisfy 
\begin{equation}
\int_{0}^{\infty}4{\pi}r^2{\omega_\rho}dr = 1.
\end{equation}
For a 3D computational domain, a widely-used form of $\omega_\rho$ is given as follows \cite{warren2003vapor}:
\begin{equation}
\omega_\rho = \frac{15}{2{\pi}r_d^3}(1 - r/r_d)^2.
\end{equation}
Notice that other forms of $\omega_\rho$ are also used in literature, e.g., Lucy kernel function~\cite{li2013three}. The mDPD model parameters must be individually calibrated for each form of $\omega_\rho$. The standard velocity Verlet algorithm~\cite{revenga1999boundary} is used to integrate the resulting equations of motion in mDPD simulations.

\section{\label{sec:model-calibration}Methodology of model calibration}

In the mDPD formulation, the thermodynamic properties of fluids are dependent only on the conservative force term in \autoref{eq:conservative-force}. To model a specific type of fluid, four model parameters from the conservative force term need to be calibrated, including the force coefficients $A$ and $B$ and cut-off radii $r_C$ and $r_d$. Most DPD systems are designed to reserve only the bulk fluid properties of key interest and neglect atomistic-scale details such as interactions and self-motion of fluid molecules. So far as the hydrocarbons residing in the environment of mesoscale (i.e. nano- to micro-scale) geological confinement are concerned, the fluid pressure can vary across a wide range of scales and mDPD is thus expected to accurately predict the fluid response in density. Meanwhile, the flow properties of hydrocarbons in mesoscale confinement can be heavily influenced by the fluid-wall interactions. Accurate modeling of free surface tension of hydrocarbons is considered an important prerequisite for the development of an mDPD fluid-wall interaction model. Therefore, the pressure-density dependency and free surface tension are chosen as two closures for the calibration of mDPD model parameters. Since the number of parameters to determine is larger than that of the closures, i.e. 4 versus 2, the calibration process is essentially an under-determined system. This indicates that multiple combinations of the parameters are possible in the solution space to satisfy the closures. In our study, we have found it practically possible to decrease the number of parameters needed to calibrate by adopting a fixed relation of $r_d = 0.75r_C$ and retain only $A$, $B$ and $r_C$ as three independent parameters.

\begin{figure}[ht]
    \centering
    \includegraphics[width=\textwidth]{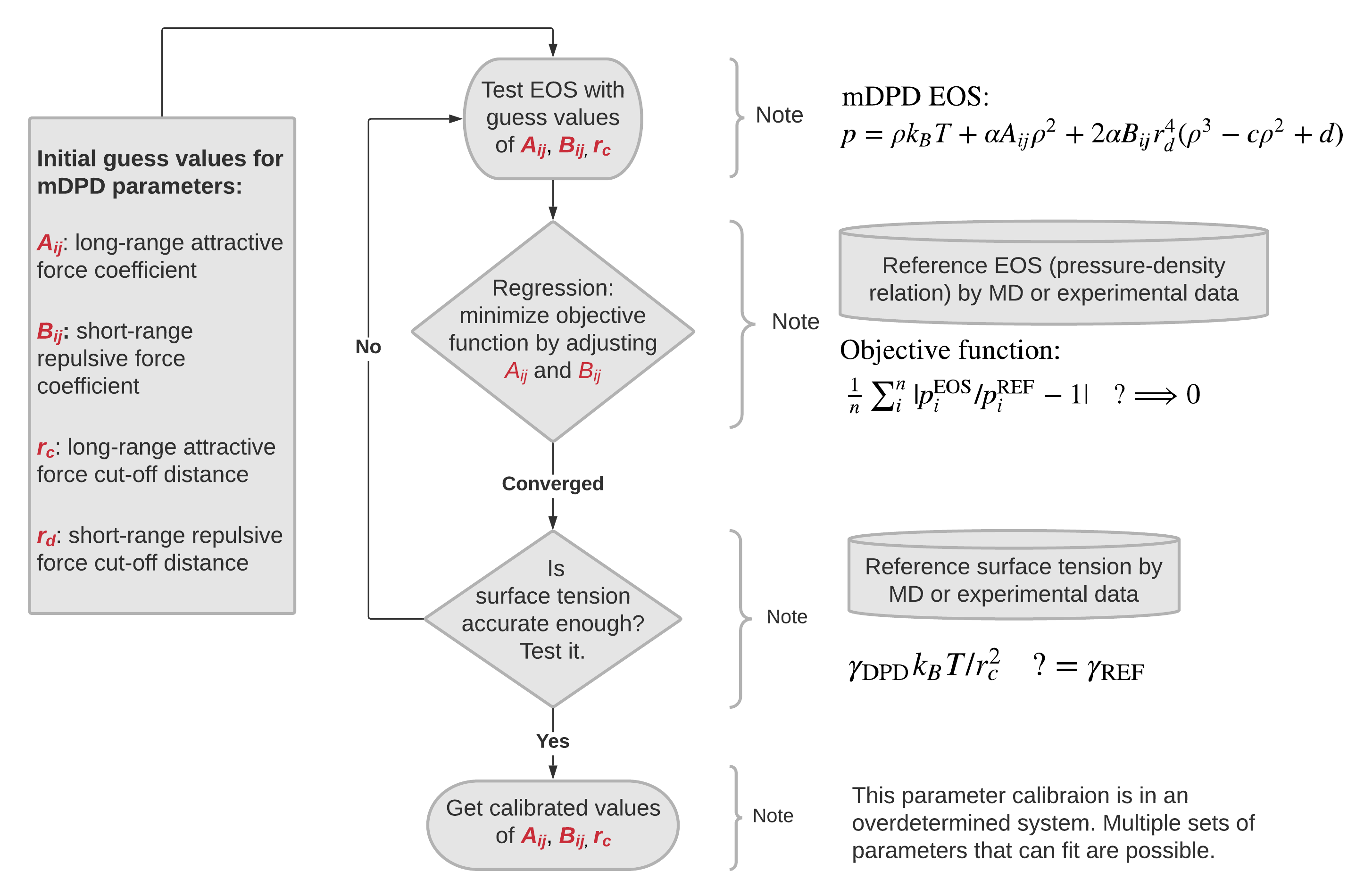}
    \caption{Schematic of a model parameter calibration process for mDPD.}
    \label{fig:DPD_upscaling}
\end{figure}

An iterative calibration process for determining the mDPD model parameters is concisely illustrated in \autoref{fig:DPD_upscaling}. The process can be concluded in two major steps.
For the first step, an isothermal equation of state (EOS) that describes the pressure-density profile for mDPD fluids is used to calibrate the $A$ and $B$ with the reference profile until a specified agreement criterion is reached. The reference data can be from experimental measurements upon the availability or molecular simulations. To begin with the calibration, we need to provide an initial guess for the three independent parameters. They can be the parameters for water \cite{ghoufi2011mesoscale} from our experience. The EOS for mDPD\cite{warren2003vapor} reads:
\begin{equation}
\label{eq:eos-function}
p = {\rho}k_BT + {\alpha}A\rho^2 + 2{\alpha}Br_d^4(\rho^3-c\rho^2+d),
\end{equation}
where $A$ and $B$ are nothing but the two parameters to calibrate, and $\alpha$, $c$ and $d$ are three fitting coefficients. Before calibrating $A$ and $B$, we examined the values of $\alpha$, $c$ and $d$ given in \citet{warren2003vapor}. We found that in order to well match the mDPD EOS profile with that calculated from the mDPD simulations, the reference fitting coefficients need to be adjusted. A combination of $\alpha = 0.101$, $c = 4.16$, $d = 19$ was found to work for the mDPD EOS for heptane, in which the values for $\alpha$ and $c$ are identical to those in \citet{warren2003vapor}, whereas $d$ is slightly modified. To determine $A$ and $B$, we define an objective function as
\begin{equation}
f = \frac{1}{N}\sum_{n=1}^{N}|p_i^{\rm EOS} / p_i^{\rm REF}-1|,
\end{equation}
where $p_i^{\rm REF}$ is the $i$-th data point of reference pressure (with a total number of data points to be $N$), and $p_i^{\rm EOS}$ is the corresponding pressure calculated from the EOS with the same density as the $i$-th reference data point. By minimizing the objective function via a nonlinear regression algorithm, a pair of $A$ and $B$ values can be obtained. For the second step, an mDPD simulation with the $A$ and $B$ obtained in the first step is performed to calculate the surface tension of an unconfined bulk fluid and check against the reference data via a mapping relationship from the DPD reduced unit to the real unit. If the simulated surface tension does not match the reference value, the $r_C$ will be adjusted and then the $A$ and $B$ will be recalibrated in the first step. The two steps are repeated until the reference pressure-density profile and free surface tension are both matched by mDPD simulations.

\section{\label{sec:results}Results and discussions}

Heptane is an alkane with an intermediate chain length and is abundant in shale oil and gas resources. The accurate and efficient modeling of bulk heptane with mDPD will serve as an important example for the development of mDPD models for other hydrocarbon fluids. Notice that this work only considers the liquid phase of heptane in the pressure-temperature conditions relevant to the natural confinement in subsurface.

\subsection{Simulation setup}

\begin{figure}[ht]
    \centering
    \includegraphics[width=0.4\textwidth]{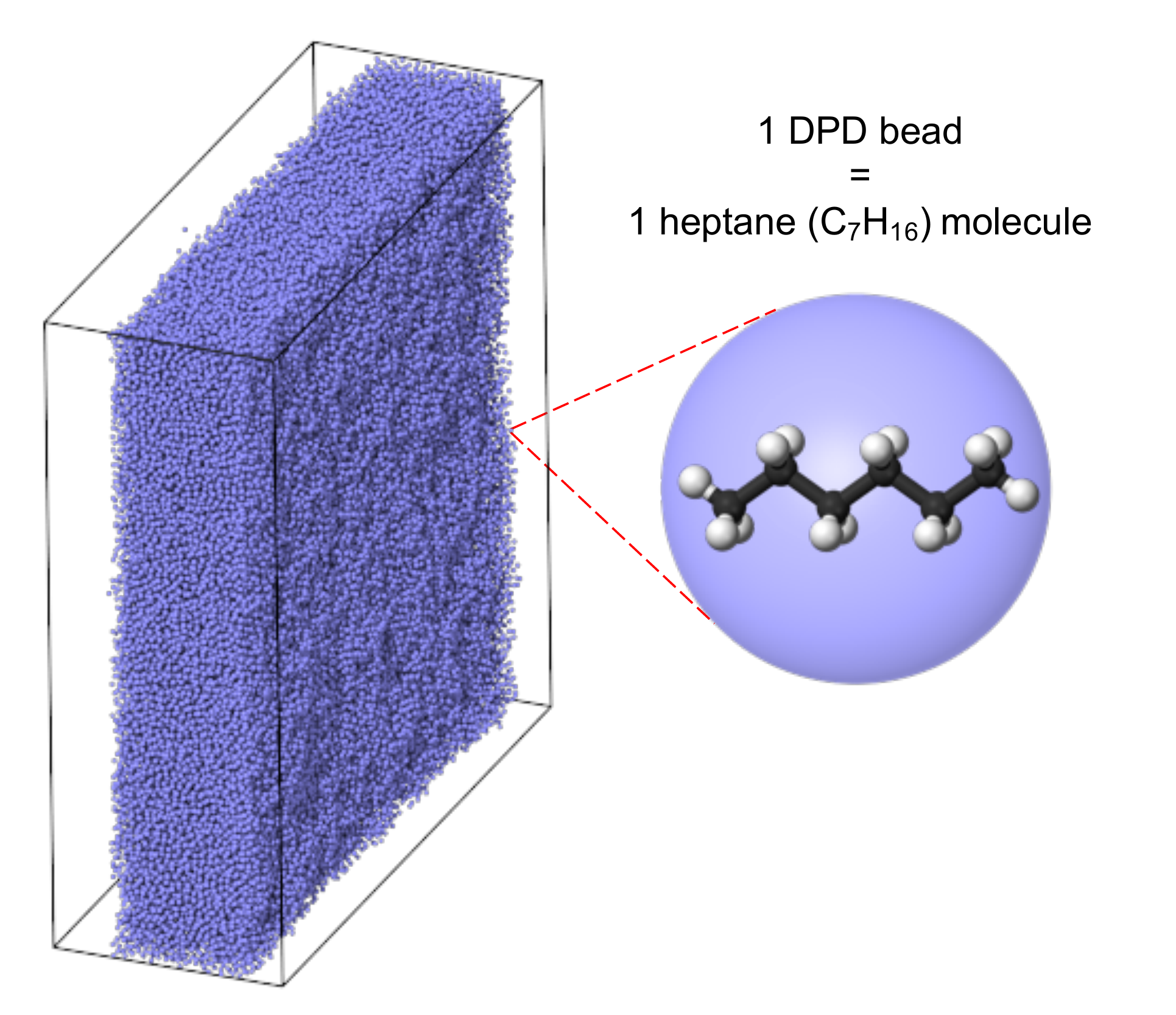}
    \caption{Simulation snapshot for the free surface of a bulk liquid modeled by the mDPD method. The bulk liquid comprises 150,000 mDPD particles.}
    \label{fig:DPD_interface}
\end{figure}

The reduced units are commonly used for DPD model parameters. The first parameter that we need to specify is the DPD coarse-graining factor, $N_{\rm m}$. In this work, $N_{\rm m}$ was chosen to be 1 for heptane, meaning that one DPD bead is used to represent the equivalent envelope volume of a heptane molecule; see \autoref{fig:DPD_interface}.
In general, $N_{\rm m}$ can be 2, 3, or higher for fluids. For example, $N_{\rm m} =3 $ is widely used for water~\cite{ghoufi2011mesoscale}. However, there is conceivably an upper limit of $N_m$ to guarantee the numerical accuracy in simulations. When $N_m$ exceeds the limit, the DPD beads will aggregate in a crystallized structure and cannot flow in simulations~\cite{pivkin2006coarse}.

To calibrate the mDPD model for the fluid pressure-density profile, we performed the mDPD simulations of 5765 - 6400 DPD beads in a periodic box with the lengths of sides, \textit{L\textsubscript{x}} = \textit{L\textsubscript{y}} = \textit{L\textsubscript{z}} = 10, in the reduced unit. To provide numerical validation reference, we conducted the NVT MD simulations of 5765 - 6400 heptane molecules in a periodic box with the lengths of sides, \textit{L\textsubscript{x}} = \textit{L\textsubscript{y}} = \textit{L\textsubscript{z}} = 112.4 \r{A}. The TraPPE-UA \cite{martin1998transferable} and OPLS-UA force fields \cite{jorgensen1984optimized} were used for heptane molecules in MD simulations. Notice that the mDPD simulation box is actually the same size with its MD counterpart after unit conversion, which will be elaborated in the following section. Further, to calibrate the mDPD model for the free surface tension of heptane, a total of 150,000 mDPD beads were used in a periodic simulation box with the lengths of sides, \textit{L\textsubscript{x}} = \textit{L\textsubscript{y}} = 50 and \textit{L\textsubscript{z}} = 20, in the reduced unit, as shown in \autoref{fig:DPD_interface}. The surface tension was calculated using the following equation:
\begin{equation}
    \gamma = \int[p_{zz} - \frac{1}{2}(p_{xx} + p_{yy})]dz,
\end{equation}
where $p_{xx}$, $p_{yy}$ and $p_{zz}$ are the diagonal components of the Cauchy stress tensor. The MD simulations were also performed with a bulk of 96,000 heptane molecules to provide numerical reference of free surface tension. The MD simulations were carried out in LAMMPS \cite{plimpton1995fast} and the mDPD simulations were conducted with an mDPD package based on LAMMPS~\cite{xia2017many}.

\subsection{Heptane at 303 K}

\autoref{tab:correspondence-303K} lists the parameters used in the reduced units and their conversion in the physical units for heptane at the ambient temperature of $T = 303$ K.
\begin{table}[ht]
\caption{Conversion of the DPD fluid properties from the reduced units to the physical units for bulk heptane at 303 K.}
\label{tab:correspondence-303K}
\centering
\begin{tabular}{ccccc}
\toprule
\multicolumn{2}{c}{DPD}& &\multicolumn{2}{c}{Physical units}\\
\cline{1-2} \cline{4-5}
Parameter & Value & DPD $\,\to\,$ real units & Parameter & Value\\
\midrule
Bead & $1$ & $N_m$ & $1$ & $1$ heptane\\
$r_c^*$ & $1$ & $(\rho^*N_mV)^{1/3}$ & $r_c$ & $11.24$ \r{A}  \\
$\rho^*$ & $5.8$ & $\rho^*N_mM/N_ar_c^3$ & $\rho$ & $675.4$ kg/m$^3$ \\
$p^*$ & $0.03$ & $p^*k_BT/r_c^3$ & $p$ & $1$ bar \\
$\gamma^*$ & $5.9$ & $\gamma^*k_BT/r_c^2$ & $\gamma$ & $19.49$ mN m$^{-1}$ \\
$\kappa^{-1*}$ & $55.8$ & $N_m/(nk_BT\kappa^{-1*})$ & $\kappa^{-1}$ & $1.06\times10^{-9}$ Pa$^{-1}$ \\
${\delta}t^{*}$ & $0.01$ & $N_{m}D_{\rm bead}r_c^3/D_{\rm heptane}$ & ${\Delta}t$ & $23.46$ ps\\
\bottomrule
\end{tabular}
\end{table}
In the table, $r_c$, $\rho$, $p$, $\gamma$, $\kappa^{-1}$, and ${\Delta}t$ are the cut-off radius, number density, pressure, surface tension, compressibility, and timestep size, respectively. $V = 246.3$ \r{A}$^3$ is the envelope volume of a heptane molecule, $M = 100.2$ ${\rm g}\cdot{\rm mol}^{-1}$ is the molar weight of a heptane molecule, $N_A$ is the Avogadro constant, $k_B$ is the Boltzmann constant, $n$ is the number of heptane molecules in the volume of $1$ m$^3$. $D_{\rm heptane}$ = $3.22 \times 10^{-9}$ m$^2$ s$^{-1}$ is the self-diffusion coefficient of heptane \cite{moore1974diffusion}, while $D_{\rm bead}$ is its counterpart in the reduced unit. $D_{\rm bead}$ is calculated from the Einstein relation:
\begin{equation}
\label{eq:einstein-relation}
{\langle}{\vert}{r(t)-r(0)}{\vert}^2\rangle\ = 2dD_{\rm bead}t,
\end{equation}
in which the term on the left stands for the mean-square displacement (MSD), $r(t)$ is the position of the bead at simulation time $t$, and $d = 3$ is the dimension of space.

\begin{figure}[ht]
    \centering
    \includegraphics[width=0.6\textwidth]{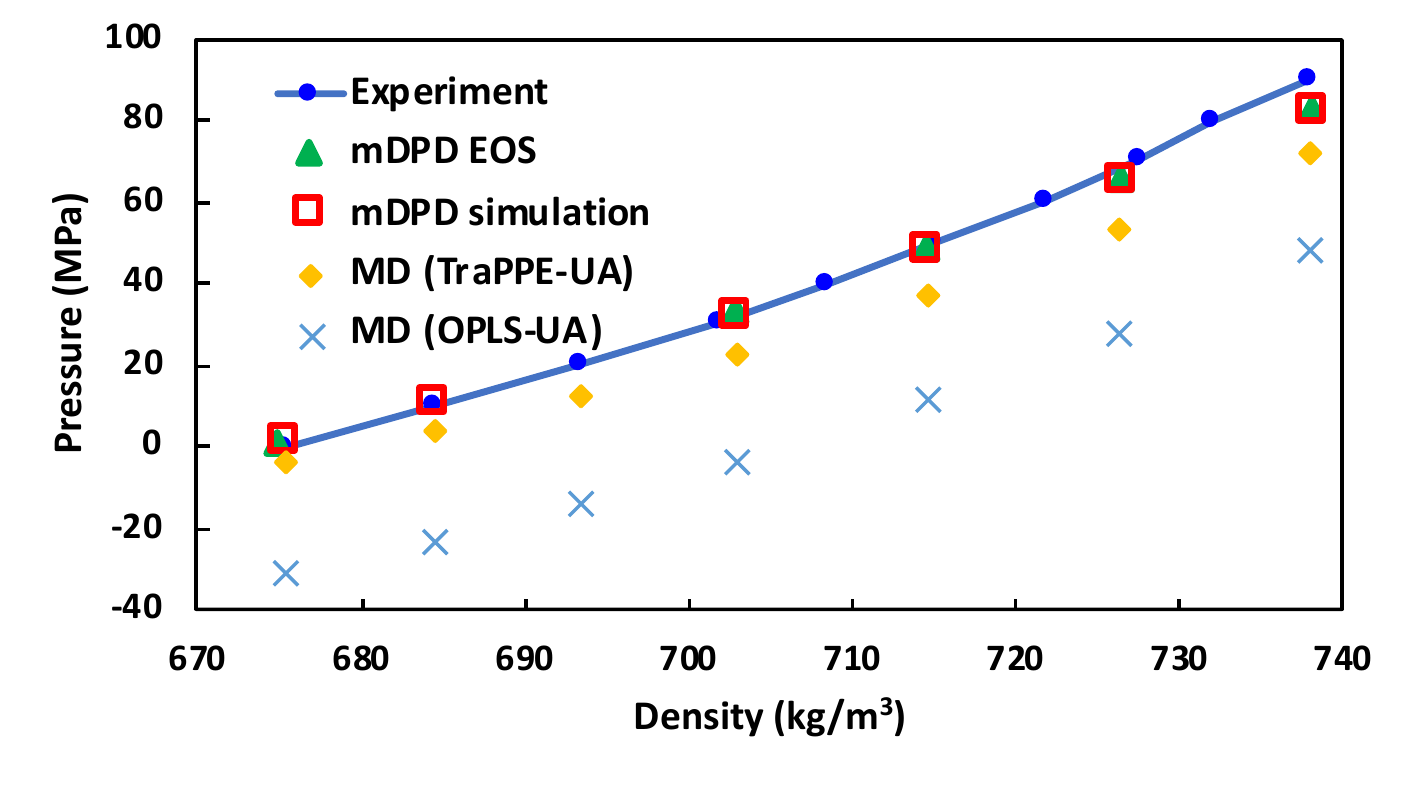}
    \caption{The fluid pressure-density profiles for bulk heptane at 303 K.}
    \label{fig:DPD_303}
\end{figure}

For $T = 303$ K, we used the experimentally-measured pressure-density dependency \cite{takaishi1998measurements} and surface tension \cite{rolo2002surface} as the primary reference data in the parameter calibration. Applying the process described in the previous section, we obtained a set of the independent parameters that satisfy the two closures, i.e. $A = -36$, $B = 25$, and $r_C = 11.24$ \r{A}. When experimental measurements are not available, MD simulations can be performed to generate reference data. In the present study, the equivalent MD simulations were also conducted to serve as additional validation. \autoref{fig:DPD_303} shows a comparison of the bulk pressure-density profiles obtained by the experimental measurements, mDPD EOS, mDPD simulations and MD simulations, respectively.
It is remarkable that the mDPD EOS and mDPD simulation results agree closely with the experimental data in the tested pressure range (0 - 100 MPa). Meanwhile, the MD simulation results obtained with the TraPPE-UA force field are slightly less accurate than those of mDPD, whereas the MD model based on the OPLS-UA force field rendered a substantial deviation from all the others. Compared with MD, the mDPD model demonstrated a high fidelity for modeling heptane. Further, our study suggests that the mDPD model can satisfy a much larger range of fluid pressure-density profiles, given further calibration of the model parameters. However, it is not necessary to make the mDPD model accurately predict the fluid density in an excessive pressure range (e.g. 100 - 200 MPa), which rarely exists in natural confinements. As a relevant note, it is worth mentioning that the difficulty to determine the model parameters can increase drastically for satisfying an extended range of fluid pressure-density profile while attempting to guarantee an accurate surface tension. 

\begin{figure}[ht]
\centering
\includegraphics[width=0.6\textwidth]{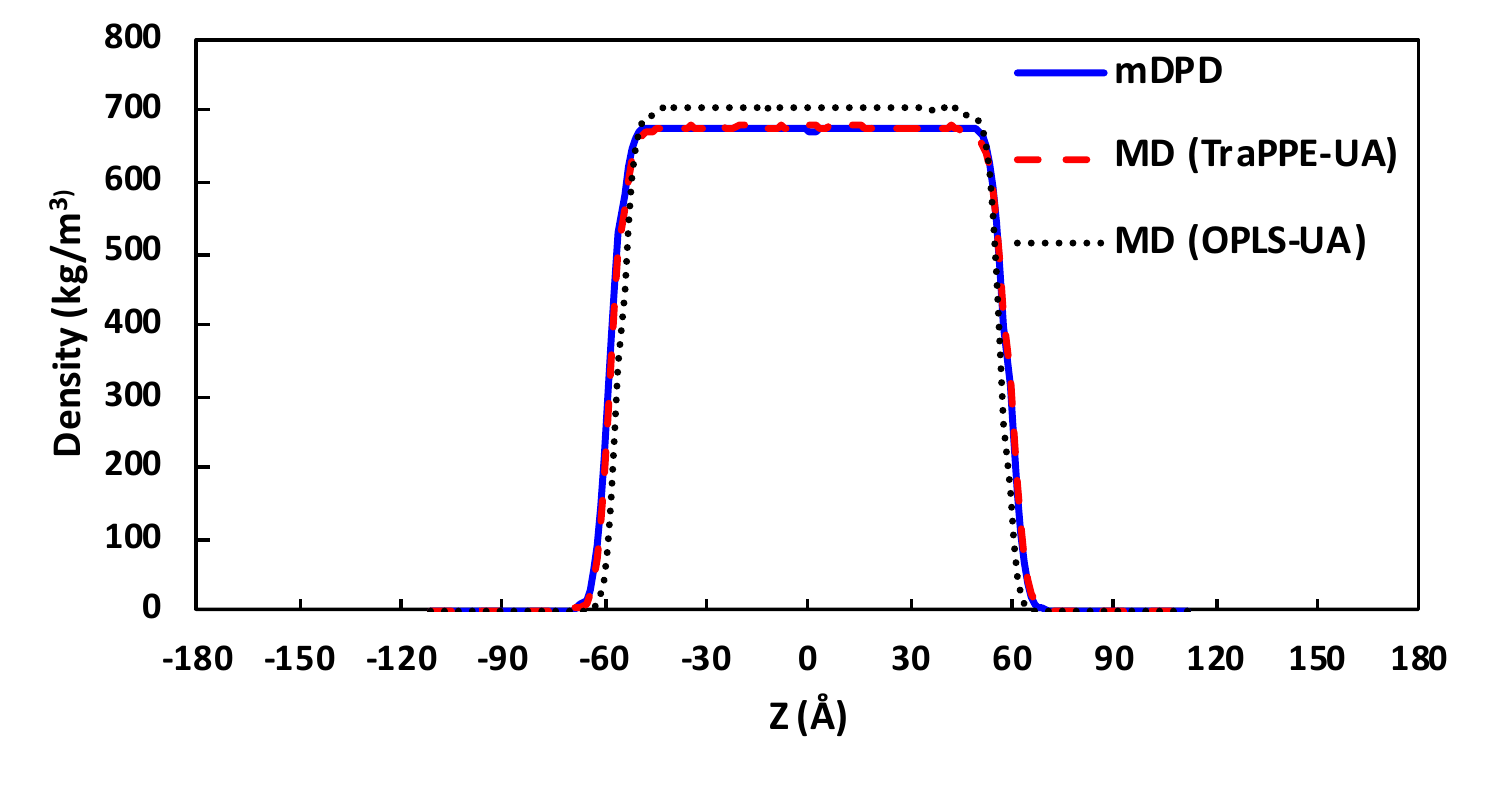}
\caption{Density profiles of heptane as a function of $z$ at 303 K.}
\label{fig:density_303}
\end{figure}

\autoref{tab:result-fitting-303} lists the surface tensions and unconfined fluid densities obtained by the experimental measurement, mDPD simulation and MD simulation, respectively, at $T = 303$K. The surface tension calculated in the mDPD simulation agrees well with the experimental value. In comparison, the TraPPE-UA based MD simulation resulted in a slight under-prediction, whereas its OPLS-UA counterpart rendered a large over-prediction. Besides, the amplitude of estimate errors in the MD simulations is one order of magnitude higher than that in the mDPD simulation, likely due to the much smaller system sizes used in the case of MD. Moreover, a comparison between mDPD and MD on the time-averaged fluid density profiles across the bulk are shown in \autoref{fig:density_303}. To obtain the density profile in the mDPD simulation, we set the bin size to be 0.1 and sampled the profile every time step over a total of one million time steps. The density profiles calculated by the mDPD and TraPPE-UA based MD simulations are close to each other and agree reasonably with the experimental value in \autoref{tab:result-fitting-303}, whereas the UPLS-UA based MD simulation shows a substantial over-prediction. Another remarkable observation is that the oscillations of fluid density near the free surface are tiny in the present mDPD simulation for heptane, whereas such oscillations were much stronger in the case of water \cite{ghoufi2011mesoscale}. The attractive force parameter $A$ is speculated as a main impact factor in the mDPD model for the oscillations at the vicinity of surface. In addition, the bulk size of fluid is evidently another factor. Compared with the bulk water simulation that comprised of 1,000 - 10,000 beads in literature \cite{ghoufi2011mesoscale}, our simulation of heptane used a much larger bulk that contained 150,000 beads. Above all, our rigorous evaluation proves an excellent fidelity of the mDPD model for predicting the key properties of bulk heptane at the ambient temperature.

\begin{table}[ht]
\caption{Heptane surface tensions and fluid densities obtained by experiments, mDPD and MD at 303K.}
\label{tab:result-fitting-303}
\centering
\begin{tabular}{ccc}
\toprule
Method &  $\gamma$ (mN m$^{-1}$) & $\rho$ (kg m$^{-3}$)\\
\midrule
Experiment  & $19.49$ \cite{rolo2002surface} & $675.4$ \cite{takaishi1998measurements}\\
mDPD  & $19.20 \pm 0.13$ & $674.0 \pm 0.3$ \\
MD (TraPPE-UA)  & $18.58 \pm 1.66$ & $677.0 \pm 1.4$ \\
MD (OPLSE-UA) & $25.19 \pm 1.87$ & $703.9 \pm 1.4$ \\
\bottomrule
\end{tabular}
\end{table}

\subsection{Heptane at 323 K}

We performed the mDPD model calibration for heptane at $T = 323$K as another example, since the experimental measurements are also are available for reference. \autoref{tab:correspondence-323K} lists the parameters used in the reduced units and their conversion in the physical units. The envelope volume for one heptane molecule, $V$, is 252.5 \r{A}$^3$, which is slightly larger than that at $T = 303$ K due to thermal expansion. The self-diffusion coefficient of heptane, $D_{\rm heptane}$ is equal to $4.47 \times 10^{-9}$ m$^2$ s$^{-1}$ \cite{iwahashi1990dynamical}, which is slightly increased compared with that at $T = 303$ K.

\begin{table}[ht]
\caption{Conversion of the DPD fluid properties from the reduced units to the physical units for bulk heptane at 323 K.}
\label{tab:correspondence-323K}
\centering
\begin{tabular}{ c c c c c }
\toprule
\multicolumn{2}{ c }{DPD}& &\multicolumn{2}{ c }{Physical units}\\
\cline{1-2} \cline{4-5}
Parameter & Value & mDPD $\,\to\,$ real units & Parameter & Value\\
\midrule
Bead & $1$ & $N_m$ & $1$ & $1$ heptane\\
$r_c^*$ & $1$ & $(\rho^*N_mV)^{1/3}$ & $r_c$ & $11.22$ \r{A}  \\
$\rho^*$ & $5.6$ & $\rho^*N_mM/N_ar_c^3$ & $\rho$ & $658.9$ kg/m$^3$ \\
$p^*$ & $0.03$ & $p^*k_BT/r_c^3$ & $p$ & $1$ bar \\
$\gamma^*$ & $4.9$ & $\gamma^*k_BT/r_c^2$ & $\gamma$ & $17.44$ mN m$^{-1}$ \\
$\kappa^{-1*}$ & $48.0$ & $N_m/(nk_BT\kappa^{-1*})$ & $\kappa^{-1}$ & $1.18\times10^{-9}$ Pa$^{-1}$ \\
${\delta}t^{*}$ & $0.01$ & $N_{m}D_{\rm bead}r_c^3/D_{\rm heptane}$ & ${\Delta}t$ & $17.90$ ps\\
\bottomrule
\end{tabular}
\end{table}

\begin{figure}[ht]
\centering
\includegraphics[width=0.6\textwidth]{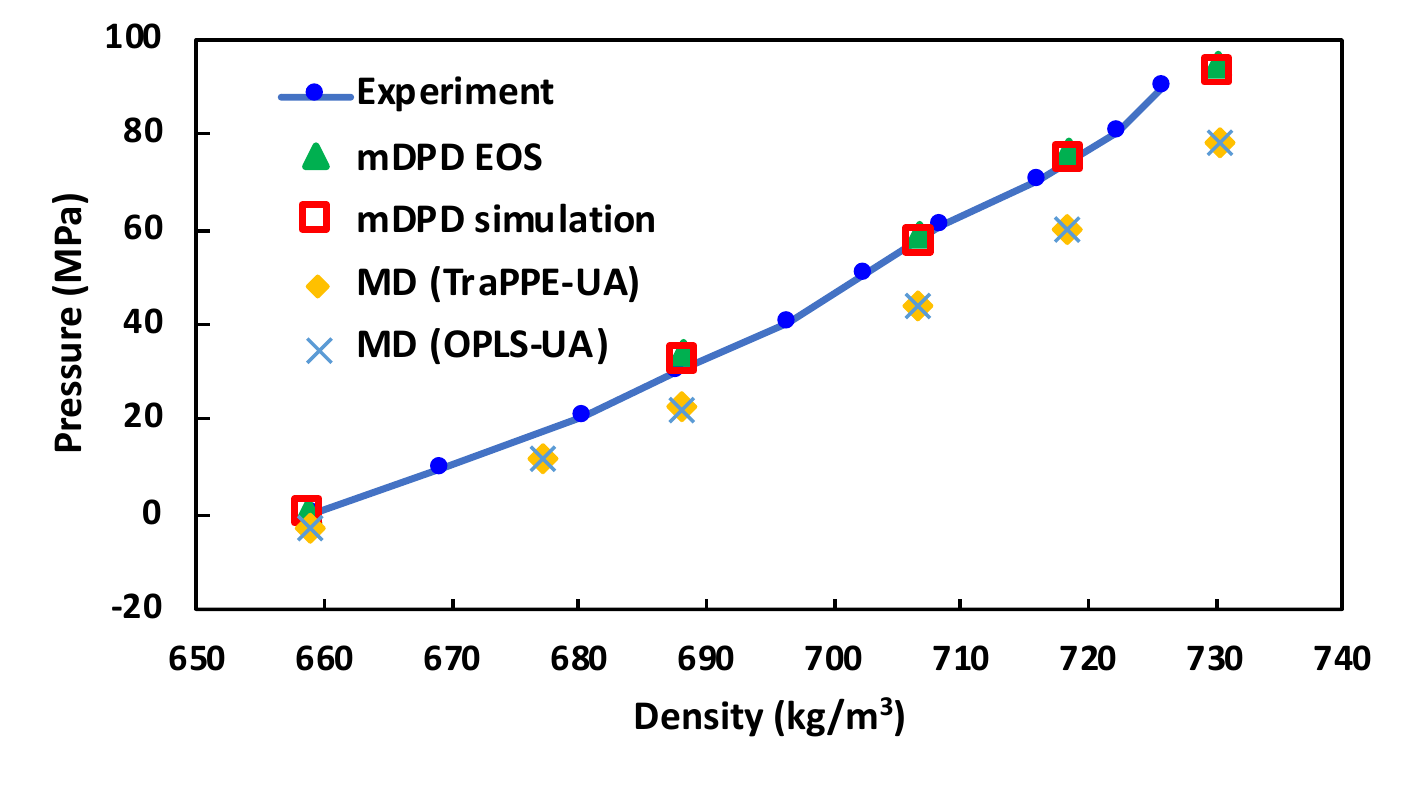}
    \caption{The fluid pressure-density profiles for bulk heptane at 323 K.}
\label{fig:DPD_323}
\end{figure}

\begin{figure}[ht]
\centering
\includegraphics[width=0.6\textwidth]{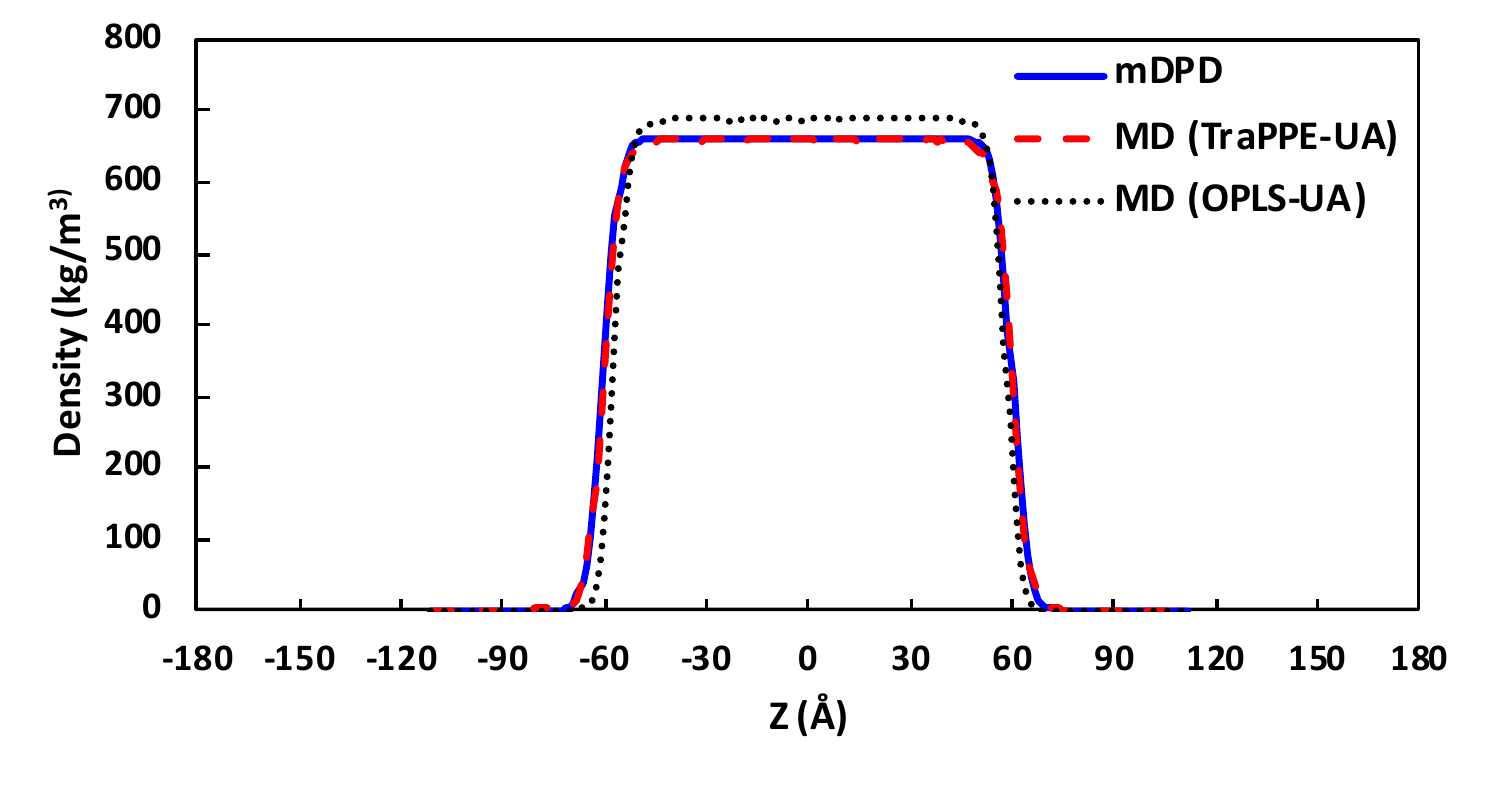}
\caption{Density profiles of heptane as a function of $z$ at 323 K.}
\label{fig:density_323}
\end{figure}

With the same calibration process, we obtained a set of the independent parameters that satisfy the two closures at $T =323$K, i.e. $A = -34$, $B = 25$ and $r_C = 11.22$ \r{A}. \autoref{fig:DPD_323} displays a comparison of the bulk pressure-density profiles obtained by the experimental measurements, mDPD EOS, mDPD simulations and MD simulations, respectively. Again, the profiles predicted by the mDPD simulations and mDPD EOS agree well with the experimental data \cite{takaishi1998measurements}. The profiles predicted by the MD simulation results based on TraPPE-UA and OPLS-UA agree with each, but rendered a slight under-prediction of the fluid pressure throughout the tested pressure range (0 - 100 MPa). \autoref{tab:result-fitting-323} lists the surface tensions and unconfined fluid density obtained by the experiments, mDPD simulations and MD simulations, respectively. Further, a comparison between mDPD and MD on the time-averaged fluid density profiles across the bulk are shown in \autoref{fig:density_323}. To obtain the density profile in the mDPD simulation, we specified the bin size to be 0.1 and sampled the profile every time step over a total of one million time steps, which were identical to the setup in the case of $T = 303$K.  Again, the mDPD model provided an accurate prediction of the surface tension and unconfined fluid density with low estimate errors, whereas the two MD models rendered either slight over-prediction or under-prediction of those properties with much higher estimate errors.  

\begin{table}[ht]
\caption{Heptane surface tensions and fluid densities obtained by experiments, mDPD and MD at 323K.}
\label{tab:result-fitting-323}
\centering
\begin{tabular}{ c c c }
\toprule
Method & $\gamma$ (mN m$^{-1}$) & $\rho$ (kg m$^{-3}$)\\
\midrule
Experiment & $17.44$ \cite{rolo2002surface}& $658.9$ \cite{takaishi1998measurements}\\
mDPD & $17.73 \pm 0.14$ & $658.6 \pm 0.1$ \\
MD (TraPPE-UA) & $16.62 \pm 1.64$ & $659.0 \pm 1.1$ \\
MD (OPLSE-UA) & $22.97 \pm 1.61$ & $688.6 \pm 1.3$ \\
\bottomrule
\end{tabular}
\end{table}

\subsection{Sensitivity of bulk size in MD}

As a side effort, we investigated the influence of bulk size on the surface tension calculation in MD simulations. At each of the two tested temperature (303 K and 323 K) and for each of the two force field models (TraPPE-UA and OPLS-UA), three simulations were conducted with the the bulk size successively quadrupled by doubling the extension of the two directions in the plane normal to the free surface, respectively, as shown in \autoref{tab:surface_tension_error}. Test results show that the increase of bulk size had little influence on the time-averaged value of surface tension. However, it is remarkable to see that the estimate error in each simulation decreased about by half over the previous one.

\begin{table}[ht]
\caption{Values of surface tension from MD with different simulation sizes at 303K and 323 K.}
\label{tab:surface_tension_error}
\centering
\begin{tabular}{ c c c c }
\toprule
T (K) & Heptane molecules & $\gamma$ (mN m$^{-1}$) (TraPPE-UA) & $\gamma$ (mN m$^{-1}$) (OPLS-UA)\\
\midrule
$303$ & $6000$ & $18.57 \pm 6.54$ & $25.24 \pm 7.16$\\
$303$ & $24000$ & $18.98 \pm 3.15$ & $25.16 \pm 3.49$\\
$303$ & $96000$ & $18.58 \pm 1.66$ & $25.19 \pm 1.87$\\
\hline
$323$ & $6000$ & $16.58 \pm 6.87$ & $22.88 \pm 6.76$\\
$323$ & $24000$ & $16.72 \pm 3.15$ & $22.75 \pm 3.33$\\
$323$ & $96000$ & $16.62 \pm 1.64$ & $22.97 \pm 1.61$\\
\bottomrule
\end{tabular}
\end{table}

\subsection{Efficiency of mDPD}

\begin{figure}[ht]
    \centering
    \includegraphics[width=0.6\textwidth]{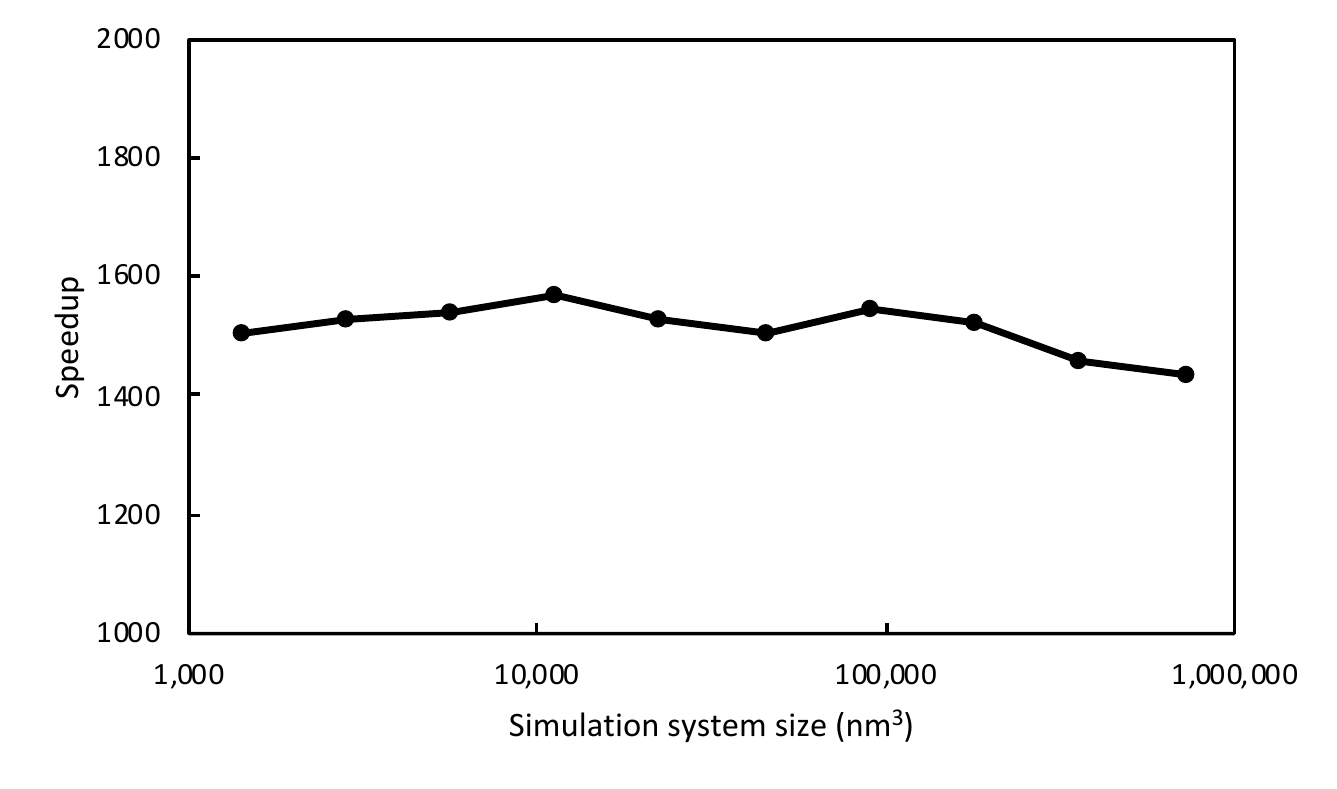}
    \caption{Speedup of mDPD over MD in equivalent simulation system volumes.}
    \label{fig:speedup}
\end{figure}

Lastly, to demonstrate the efficiency of mDPD, we performed a comparative timing test for mDPD and MD simulations in a series of equivalent unconfined volumes. All the simulations were conducted on a laptop using one CPU core (2.6 GHz Intel Core i7) for timing. This work uses the speedup of mDPD over MD as an indicator of efficiency, which is defined as the ratio of wall time for mDPD over that for MD to simulate a specified physical time duration. The test results are displayed in \autoref{fig:speedup}, showing a speedup between 1400x and 1600x for the bulk volume ranging from one thousand nm$^3$ to nearly one million nm$^3$.

\section{\label{sec:conclusions}Conclusions}

Accurate and efficient numerical models for the flow of hydrocarbons are considered essential research means in the development of enhanced geotechnical engineering for energy source recovery and carbon capture \& storage in low-porosity, low-permeability rock formations. The interconnected pore channels in which hydrocarbons may reside and flow manifest a multiscale distribution of pore aperture sizes ranging from a few nanometers to a few micrometers. A numerical model for the flow of hydrocarbons that can efficiently handle both the sub-continuum fluid dynamics in the nanoscale and continuum (or continuum-like) fluid dynamics in the microscale at the same time is much desired. To address such needs, we have developed an atomistically-validated, mesoscopic fluid flow model for heptane based on the many-body dissipative particle dynamics (mDPD) method. In this mesoscopic model, each heptane molecule is coarse-grained in one mDPD bead and the mDPD model parameters corresponding to specified temperatures have been calibrated with a rigorous upscaling approach using reference data, including experimental measurements (whenever available) and/or molecular simulations. The calibrated model parameters for heptane at two representative temperatures are summarized in \autoref{tab:result-list}.
\begin{table}[ht]
\centering
\footnotesize
\caption{The mDPD parameters calibrated for bulk heptane at 303 K and 323 K.}
\label{tab:result-list}
\begin{tabular}{ c c c c c }
T (K) & $A$ & $B$ & $r_c$ (\r{A}) & $r_d$ (\r{A})\\
\midrule
$303$ & $-36$ & $25$ & $11.24$ & $8.43$ \\
$323$ & $-34$ & $25$ & $11.22$ & $8.415$ \\
\end{tabular}
\end{table}
Our numerical results have shown that this mDPD model accurately predicts the bulk pressure-density relation of heptane as well as the free surface tension in specified temperatures. Notice that our upscaling approach can be used to calibrate the mDPD model for other hydrocarbons as well, though heptane is chosen in this work as a representative source fluid for its abundance in source rocks. The biggest advantage of mDPD is that the critical time-step sizes in the simulations are much larger than those in MD. Our timing test has demonstrated that the mDPD model is three orders of magnitude faster than its MD counterpart for simulations of bulk heptane in an equivalent volume. One limitation to the current mDPD model is that it applies to isothermal systems only. A more sophisticated mDPD model to be developed by considering total energy conservation will permit non-isothermal systems. Above all, the mDPD model development for the flow of bulk hydrocarbons in this work serves as a prerequisite for our follow-on model development for hydrocarbons confined in pore channels. The modeling of hydrocarbon flow in confinement indicates the need for considering integration of robust fluid-solid interaction models \cite{pivkin2005new,li2018dissipative,zhang2018easy} to accurately account for capillary flow \cite{arienti2011many} and slip flow \cite{xu2019new}.

\section*{Acknowledgement}

The research is primarily supported by EFRC-MUSE, an Energy Frontier Research Center funded by the U.S. Department of Energy, Office of Science, Basic Energy Sciences under Award No. DE-SC0019285.
The software development in this work is supported through the Idaho National Laboratory (INL) Laboratory Directed Research \& Development (LDRD) Program under the U.S. Department of Energy Idaho Operations Office Contract DE-AC07-05ID14517.
The research used resources in the High Performance Computing Center at INL, which is supported by the Office of Nuclear Energy of the U.S. Department of Energy and the Nuclear Science User Facilities under Contract No. DE-AC07-05ID14517.

\bibliographystyle{plainnat}

\end{document}